\documentclass[conference]{IEEEtran}

\usepackage{cite}
\usepackage{color}
\usepackage{amsmath}
\usepackage{amssymb}
\usepackage{glossaries}
\usepackage{graphicx}

\newacronym{alarp}{ALARP}{As Low As Reasonably Practicable}
\newacronym{arp}  {ARP}  {Aerospace Recommended Practice}
\newacronym{pps}  {PPS}  {Plant Protection Software}
\newacronym{rtca} {RTCA} {Requirements and Technical Concepts for Aviation}
\newacronym{sut}  {SUT}  {Software Under Test}
\newacronym{vdm}  {VDM}  {Vienna Development Method}

\newcommand{\nf}{\normalfont}
\newcommand{\ie}{\it i.e. \nf}
\newcommand{\eg}{\it e.g. \nf}

\begin{document}

\title{The Utility and Practicality of Quantifying Software Reliability}
\author{\IEEEauthorblockN{Rob Ashmore}
\IEEEauthorblockA{Defence Science and Technology Laboratory\\
Portsdown Hill Road, Fareham\\
Hants. PO17 6AD\\
Email: rdashmore@dstl.gov.uk}
}

\maketitle

\begin{abstract}
We argue that quantifying software reliability is important in demonstrating that system-level risks are \gls{alarp}. Furthermore, we demonstrate that such quantification is possible in at least one meaningful case. It is, however, unlikely to be practical in every case. This means it is unlikely to be included as an explicit objective in standards. Hence, for those cases where software reliability can be quantified, merely following a standard may lead to risk-reduction opportunities being missed.
\end{abstract}

\section{Introduction}

One way of assessing the efficacy of standards is demonstrating the efficacy of part, or parts, of a standard. We are primarily interested in the design, implementation and testing phases of software development. More specifically, we wish to quantify the probability that software will act as specified in its requirements.

\section{Qualitative Approaches} \label{sect:qualitative}

The efficacy of qualitative (\eg process-based) approaches to software development is demonstrated by the large volume of safety-critical software in use today. However, these approaches do not allow software reliability to be quantified in system fault trees. Whilst this lack of quantification is a common approach, and is consistent with \gls{arp} 4761 \cite{SAE_ARP4761}, it is not entirely benign.

\begin{figure}[!t]
\centering
\includegraphics[width=0.45\textwidth]{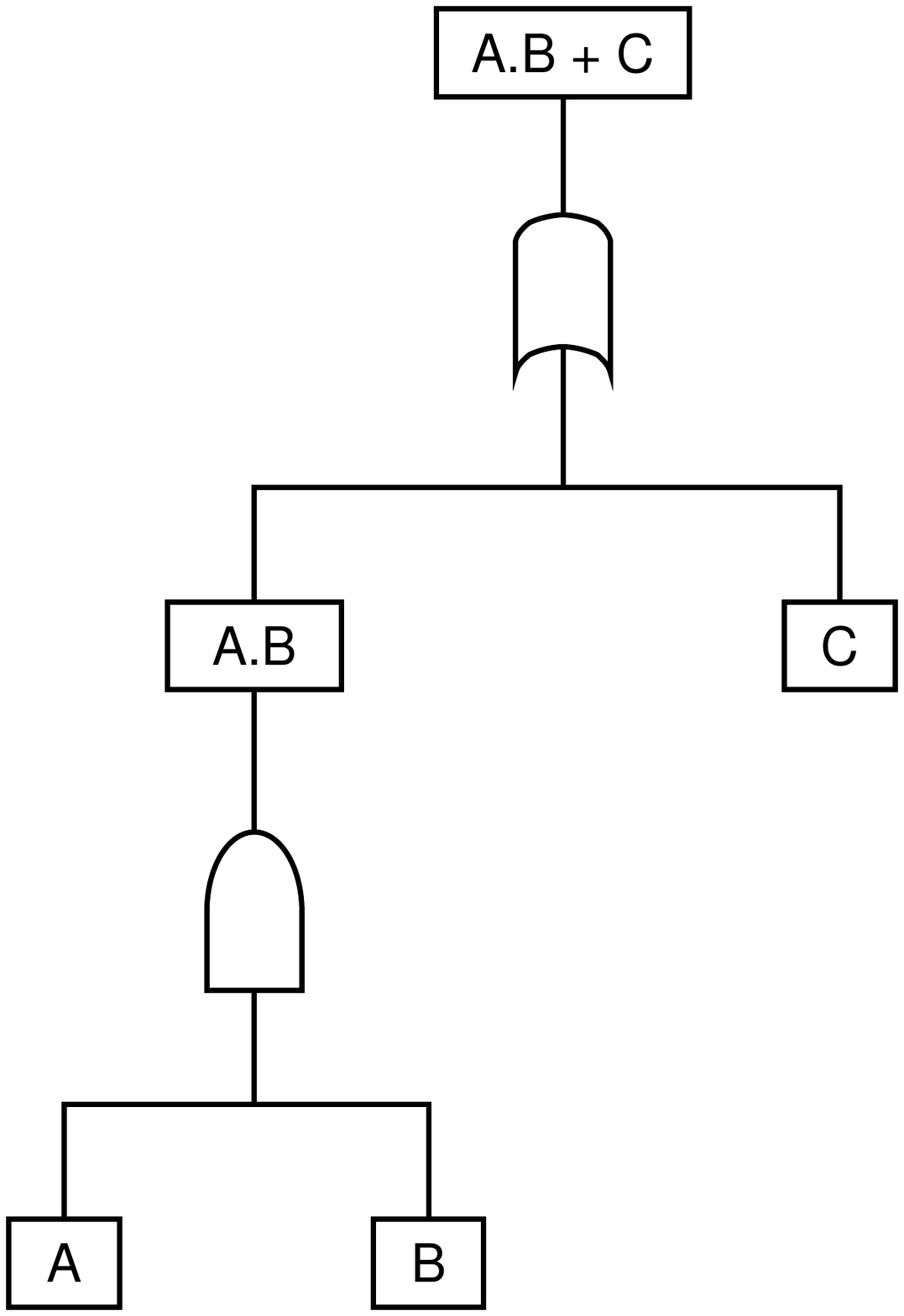}
\caption{Example Fault Tree}
\label{fig:egFaultTree}
\end{figure}

Consider the simple fault tree shown in Figure~\ref{fig:egFaultTree}. We assume that element A is implemented in software and that B and C are implemented via physical means. In this arrangement software and a mechanical item provide mutual back-up.

Suppose that the failure rates of B and C are $10^{-2}$ and $10^{-5}$, respectively. One common approach, involves giving A, the software component, an implausibly low failure rate (\eg $10^{-31}$). This rate dominates the \tt AND\nf { }gate at the bottom left. As a result, the dominant factor into the top \tt OR\nf { }gate is C's failure rate. Any attempts to reduce the overall failure rate will begin by focussing on C.

Now, suppose that we were only able to justify a failure rate of $10^{-2}$ for the software component (\ie for A). In this case the top \tt OR\nf { }gate is dominated by the combination of A and B. Hence, improvements to the overall failure rate will initially focus on these components. 

This example is obviously highly simplified: the fault tree only includes three components and, more significantly, it is highly unlikely that decisions on which components to ``improve'' will be based solely on mathematical calculations. However, the underlying point remains valid, namely that assuming extremely high rates of software reliability can highlight the \em wrong \nf system components\footnote{For reasons of brevity we do not fully detail the argument here, but a similar argument shows that related problems arise if software is given an extremely high failure rate, which is another commonly adopted approach.}. This can lead to scarce resources, both intellectual and physical (\eg power and weight budgets), being used in a sub-optimal manner, potentially jeopardising claims that risks are \gls{alarp}. 

\section{Quantification Theories} \label{sect:quantification}

The quantification of software reliability is not new: quantified estimates of the Sizewell B \gls{pps} reliability were produced in the 1990s \cite{May_Pps}. Despite this history, it is not a widely used technique. This may be partly explained by philosophical concerns (\eg the fact that software does not ``wear out'' like physical components do), which are outside the scope of this paper. Until recently, there have also been practical barriers to demonstrating meaningful levels of software reliability for non-trivial cases. 

One simple approach to quantifying software reliability uses a ``balls and urn'' model, where individual balls represent specific inputs and the distribution of balls represents the distribution of inputs the software is expected to see in operational use. Subject to certain assumptions, successfully completing $4.61 \times 10^6$ tests is sufficient to claim a reliability of $1 \times 10^{-6}$ with 99\% confidence \cite{Bishop_TowardFormalism}. 

Our point is not that great faith should be put in the precise accuracy of these numbers; like all values used in fault trees there will inevitably be some uncertainty and inaccuracy. However, simply being able to estimate software reliability to within, say, an order of magnitude will allow meaningful system-level decisions to be made.

Other approaches to quantifying software reliability may also be feasible. Consider a piece of monitoring software that initiates ``make safe'' actions under certain conditions. If the number of different inputs that meet these conditions is suitably small (\eg millions) then all of them may be explicitly tested. This would allow a numerical value to be assigned to a part of a fault tree that considered something like: ``software fails to request make safe when conditions require it''. Of course, any safety argument that used this ``partial exhaustive'' testing would also need to consider the implications of ``make safe'' being enacted when it was not required. 

In a small number of cases it may be possible to test the full input domain of the software. We have recently implemented an automated test environment, which was used to analyse (amongst other things) a currently-fielded control algorithm. Even a naive implementation of the test environment, running on commodity hardware, achieved 320 test cases per second per core\footnote{In this case most time is spent running the test oracle, which is an animated \gls{vdm} specification. Running just the \gls{sut} we can achieve around a million cases per second per core.}. This suggests that within 24 hours a modest 32-core system would be capable of exhaustively testing software that takes a combination of three 8-bit integers and three 1-bit flags. 

Our experiences show that exhaustive testing of meaningful algorithms is now within reach. Of course, it is trivially easy to produce algorithms that cannot be exhaustively tested. However, we conjecture that, even if exhaustive testing is impossible, many examples of embedded software are sufficiently small to allow millions of tests to be completed easily. This opens up the possibility of providing evidence to support statistical claims regarding software reliability.

\section{Threats to Validity} \label{sect:threats}

Our focus on software testing does not cover the case where the original requirements may be incorrect. Likewise, we have ignored interactions between software and people.

We have also implicitly assumed that the test hardware faithfully replicates the operational hardware, which is one of the assumptions underpinning the relationship between the number of successful tests and a quantified reliability claim: recent work on emulation technologies may help demonstrate that the required faithful replication has been achieved. 

Another assumption underpinning the relationship between estimated reliability and the number of successfully completed tests is that the distribution of test cases accurately reflects the software's operational input distribution. Estimates of this distribution may be informed by: discussions with stakeholders \cite{Musa_OpProfile}; development activities; and in-service data. Several different distributions could even be used. Whilst this would increase the number of tests it would not reach an impractical level.

Automatically conducting large numbers of tests needs an automated way of checking the result: that is, we need a test oracle. In some cases this may be easy to produce (\eg checking a ``make safe'' flag, or animating a formal specification). In others it may be more difficult to capture the \em entire \nf set of requirements in an oracle. However, even in these cases testing could confirm, for example, that function pre- and post-conditions are highly likely to be satisfied. This could inform key parts of system fault trees.

There are types of software that by their very nature pose particular problems to quantifying reliability. The lack of knowledge of internal state information means ``black box'' software is one example; real-time software is another \cite{Butler_Infeasible}. Adapting the ``balls and urn'' model indicates that, for example, 4.61 million hours of testing would be required. (The move from ``number of tests'' to ``hours of testing'' reflects the desire to talk about ``probability of failure per hour'' rather than ``probability of failure per demand''.) Testing for millions of hours is stretching the bounds of plausibility. That said, if the software is performing a monitoring function then, as discussed earlier, it may be possible to do a partial exhaustive test on those input sets that require action.

\section{Conclusion} \label{sect:conclusion}

The approaches to quantifying software reliability discussed above will not be suitable for every piece of software. This means they are highly unlikely to be mandated in a standard. However, when they are suitable these methods provide valuable support to system-level risk management. Without this support risk-reduction opportunities may be missed, potentially jeopardising claims that risks are \gls{alarp}.

\vspace{6pt} \footnotesize Disclaimer: This article is an overview of MOD sponsored research and is released for informational purposes only. The contents of this article should not be interpreted as representing the views of the MOD, nor should it be assumed that they reflect any current or future MOD policy. The information contained in this article cannot supersede any statutory or contractual requirements or liabilities and is offered without prejudice or commitment. \nf

\vspace{6pt} \footnotesize \copyright British Crown copyright 2014 Dstl. Published with the permission of the Controller of Her Majesty’s Stationery Office \nf

\bibliographystyle{IEEEtran}
\bibliography{EDCC}

\end{document}